\documentclass[twocolumn,showpacs,aps,prb,letterpaper,nofootinbib,raggedbottom,nobalancelastpage]{revtex4-1}
\usepackage{graphicx,bm}
\usepackage{amsmath}
\usepackage{color}

\begin{document}
\title{Effect of phonon coupling on cooperative two-photon emission from two-quantum dots}
\date{\today}
\author{J. K. Verma, Harmanpreet Singh and P. K. Pathak}
\address{School of Basic Sciences,
Indian Institute of Technology Mandi, Kamand, H.P. 175005, India}
\begin{abstract}
We predict dominating cooperative two-photon emission from two quantum dots coupled with a single mode photonic crystal cavity. The cooperative two photon emission occurs when excitons in two off-resonantly coupled quantum dots decay simultaneously. The interaction with common cavity field leads to cavity induced two-photon emission which is strongly inhibited by electron phonon coupling. The interaction with common phonon bath produces phonon induced two-photon emission which increases on increasing temperature. For identical quantum dots cavity induced two-photon emission is negligible but phonon induced two-photon emission could be large.
\end{abstract}
\maketitle
\section{Introduction}Cooperative emission by an ensemble of $N$ identical two level atoms has been the subject of intense theoretical and experimental research after the discovery of supperradiance by Dicke\cite{dicke}. It has been shown that the initial state as a symmetric superposition of atomic states leads to supperradiance, whereas antisymmetric superpositions of atomic states get decoupled from the environment and radiate negligible intensity\cite{haroche}. Such interesting effects arise due to quantum interference between different possible atomic transitions\cite{twoatom}. The superradiant and subradiant behavior has also been observed in the cooperative emission of two emitters\cite{twoatom,twoemitter} in a cavity. Further, the concept of superradiance and subradiance has been extended to the inhomogeneously broadened ensembles such as densely spaced semiconductor quantum dots (QDs) coupled to a microcavity\cite{qdsuperrad}.

Recently, there have been considerable interests in developing on chip photonic circuits using QDs coupled with photonic crystal microcavities and waveguides, particularly for the purpose of scalable quantum information processing\cite{qi}. Owing to strong electrons and holes confinement, QDs have atom like discrete energy levels and with current technology it is possible now to deterministically position a QD at desired position in photonic crystal microcavities with very high accuracy\cite{hennessy}. A significant technological progress has been made in realizing these systems, ultra high quality cavities\cite{ultraq}, ultra low-loss waveguides have been designed\cite{waveguide}. Further, incoherent\cite{incoherent} as well as coherent excitation\cite{coherent} techniques, and strong coupling regime in QD-microcavity coupled systems\cite{hennessy,strongc} have been realized. Jaynes-Cummings ladder\cite{ladder}, where more than one photon interaction with a single two level system becomes significant, has also been observed. Clearly these developments have proved photonic systems as a potential candidate for developing integrated photonic technology and scalable quantum information circuits. However, most of these studies involve interaction of the single QD with electromagnetic field. In this paper, we consider cooperative two-photon emission from two separated QDs. We are particularly interested in cooperative two-photon emission from two off-resonant QDs when the probabilities of single photon emissions could be very small. Dominating two-photon emission occurs when two-photon resonant condition is satisfied. Under two-photon resonant condition, possible two-photon transitions become indistinguishable and interfere constructively.
The two-photon resonant condition can be satisfied either for two unidentical QDs having different dipole coupling constants and exciton transition frequencies or for two identical QDs having same dipole coupling constants and exciton transition frequencies. We specifically bring out the role of phonon coupling in cooperative two-photon emission from two QDs. In semiconductor cavity quantum electrodynamics coupling with phonon bath is a unique phenomenon which is primarily responsible for exciton dephasing\cite{dephasing}. Other important processes such as off-resonant cavity mode feeding\cite{modefeed,meq}, phonon mediated population inversion\cite{inversion}, phonon assisted biexciton generation\cite{biexciton} have also been observed. We notice that there have been some interesting theoretical as well as experimental results demonstrating coupling between two quantum dots induced by common interacting field\cite{3level}. In the system of two QDs, interaction with common phonon field also plays a significant role and the phonon mediated coupling between two QDs has been recently observed\cite{arka}.

Our paper is organized as follows. In Sect.II, we present our model for resonant two-photon emission and theoretical frame work using recently developed master equation techniques\cite{meq}.
The population dynamics, probabilities for photon emissions and spectrum of the generated photons is presented in Sect.III.
Finally, we conclude in Sect.IV.
\section{Two QDs interacting with a single mode Cavity}
\label{Sec:unidentical}
We consider two separated QDs embedded in a single mode photonic crystal cavity. The energy levels of ith QD are represented by $|g_i\rangle$
and $|e_i\rangle$, for $i=1,2$, corresponding to ground state and exciton state. The Hamiltonian in the rotating frame
is given by
\begin{eqnarray}
H=\hbar\delta_1\sigma_1^+\sigma_1^-+\hbar\delta_2\sigma_2^+\sigma_2^-+\hbar g_1(\sigma_1^+a+a^{\dag}\sigma_1^-)\nonumber\\
+\hbar g_2(\sigma_2^+a+a^{\dag}\sigma_2^-)+H_{ph},
\label{hm}
\end{eqnarray}
where $\sigma_i^+=|e_i\rangle\langle g_i|$, $\sigma_i^-=|g_i\rangle\langle e_i|$, $\delta_i=\omega_i-\omega_c$, $\omega_c$ is frequency of the cavity mode, $\omega_i$ is transition frequency for exciton energy level
and $g_i$ is the coupling constant for ith QD, $a$ and $a^{\dag}$ are photon annihilation and creation operators, respectively. The phonon bath and exciton phonon interaction is included in
$H_{ph}=\hbar\sum_k\omega_kb_k^{\dag}b_k+\lambda_k\sigma_1^+\sigma_1^-(b_k+b_k^{\dag})+\mu_k\sigma_2^+\sigma_2^-(b_k+b_k^{\dag})$ with $b_k(b_k^{\dag})$ as phonon annihilation (creation) operator for k-th mode.
In order to understand the influence of exciton-phonon interaction we made polaron transform. The transformed Hamiltonian
$H^{\prime}=e^PHe^{-P}$ with
$P=\sigma_1^+\sigma_1^-\sum_k\frac{\lambda_k}{\omega_k}(b_k-b_k^{\dag})+\sigma_2^+\sigma_2^-\sum_k\frac{\mu_k}{\omega_k}(b_k-b_k^{\dag})$;
is separated into cavity-QD system, phonon bath and system-bath interaction as $H^{\prime}=H_s+H_b+H_{sb}$, where
\begin{eqnarray}
 H_s=\hbar\Delta_1\sigma_1^+\sigma_1^-+\hbar\Delta_2\sigma_2^+\sigma_2^-+\langle B\rangle X_g,\\
 H_b=\hbar\sum_k\omega_k b_k^{\dag}b_k,\\
 H_{sb}=\xi_gX_g+\xi_uX_u,
\end{eqnarray}
where the polaron shifts $\sum_k\lambda_k^2/\omega_k$, $\sum_k\mu_k^2/\omega_k$ are included in the effective detunings $\Delta_1$ and $\Delta_2$.
The system operators are given by $X_g=\hbar(g_1\sigma_1^+a+g_2\sigma_2^+a)+H.c.$, $X_u=i\hbar(g_1\sigma_1^+a+g_2\sigma_2^+a)+H.c.$ and bath fluctuation operators are $\xi_g=\frac{1}{2}(B_++B_--2\langle B\rangle)$ and $\xi_u=\frac{1}{2i}(B_+-B_-)$. The phonon
displacement operators are $B_{\pm}=\exp[\pm\sum_k\frac{\lambda_k}{\omega_k}(b_k-b_k^{\dag})]=\exp[\pm\sum_k\frac{\mu_k}{\omega_k}(b_k-b_k^{\dag})]$ with expectation value $\langle B\rangle=\langle B_+\rangle=\langle B_-\rangle$. We use transformed Hamiltonian $H^{\prime}$ to derive polaron master equation for describing the dynamics of the system. After making Born-Markov
approximation, the master equation is derived in Lindblad form. The Lindblad super operator corresponding to an operator $\hat{O}$
is defined as
${\cal L}[\hat{O}]\rho=\hat{O}^{\dag}\hat{O}\rho-2\hat{O}\rho\hat{O}^{\dag}+\rho\hat{O}^{\dag}\hat{O}$.
The spontaneous emission, cavity damping and phonon induced dephasing are also included
in the master equation. The final form of master equation in terms of density matrix for cavity-QDs coupled system $\rho_s$ is written as\cite{meq}
\begin{eqnarray}
\dot{\rho_s}=-\frac{i}{\hbar}[H_s,\rho_s]-{\cal L}_{ph}\rho_s-\frac{\kappa}{2}{\cal L}[a]\rho_s\nonumber\\
-\sum_{i=1,2}\frac{\gamma_i}{2}{\cal L}[\sigma_i^-]\rho_s
-\frac{\gamma_i^{\prime}}{2}{\cal L}[\sigma_i^+\sigma_i^-]\rho_s,
 \label{meq}
\end{eqnarray}
where $\kappa$, $\gamma_i$, $\gamma^{\prime}_i$ are cavity leakage, spontaneous decay, dephasing rates, and
\begin{eqnarray}
 {\cal L}_{ph}\rho_s=\frac{1}{\hbar^2}\int_0^{\infty}d\tau\sum_{j=g,u}G_j(\tau)[X_j(t),X_j(t,\tau)\rho_s(t)]+H.c.
\end{eqnarray}
with $X_j(t,\tau)=e^{-iH_s\tau/\hbar}X_j(t)e^{iH_s\tau/\hbar}$, and polaron Green functions are given by
$G_g(\tau)=\langle B\rangle^2\{\cosh[\phi(\tau)]-1\}$ and $G_u(\tau)=\langle B\rangle^2\sinh[\phi(\tau)]$. In this master equation
system-phonon interaction is included in phonon correlation function $\phi(\tau)$. The phonon bath is treated as a continuum
with spectral function $J(\omega)=\alpha_p\omega^3\exp[-\omega^2/2\omega_b^2]$, where the parameters $\alpha_p$ and $\omega_b$ are
the electron-phonon coupling and cutoff frequency respectively. In our calculations we use $\alpha_p=1.42\times10^{-3}g_1^2$ and $\omega_b=10g_1$, which gives $\langle B\rangle=0.90$, $0.84$, and $0.73$ for $T=5K$, $10K$, and $20K$, respectively, which matches with recent experiments\cite{asymmetric,meq}. The phonon correlation function is given by
\begin{eqnarray}
\phi(\tau)=\int_0^{\infty}d\omega\frac{J(\omega)}{\omega^2}
\left[\coth\left(\frac{\hbar\omega}{2K_bT}\right)\cos(\omega\tau)-i\sin(\omega\tau)\right],
\end{eqnarray}
where $K_b$ and $T$ are Boltzmann constant and the temperature of phonon bath respectively.

We are interested in two-photon cooperative emission from two QDs, therefore we work in the condition when single photon transitions are suppressed
from individual QDs, i.e. the coupling constants of cavity field with QDs are much smaller than their detunings ($g_1,~g_2\ll\Delta_1,~\Delta_2$). Under such
condition the master equation (\ref{meq}) can be further simplified, using $H_s=\hbar\Delta_1\sigma_1^+\sigma_1^-+\hbar\Delta_2\sigma_2^+\sigma_2^-$ and neglecting the terms proportional to $g_1$ and $g_2$ in the expression of $X_j(t,\tau)$.
The simplified form of master equation provide clear picture of processes involved in the dynamics. Under such approximation the master
equation (\ref{meq}) takes the form
\begin{eqnarray}
\dot{\rho_s}=-\frac{i}{\hbar}[H_{eff},\rho_s]-\frac{\kappa}{2}{\cal L}[a]\rho_s\nonumber\\
-\sum_{i=1,2}\left(\frac{\gamma_i}{2}{\cal L}[\sigma_i^-]
+\frac{\gamma_i^{\prime}}{2}{\cal L}[\sigma_i^+\sigma_i^-]
+\frac{\Gamma^+_i}{2}{\cal L}[\sigma_i^+a]+\frac{\Gamma^-_i}{2}{\cal L}[a^{\dag}\sigma_i^-]\right)\rho_s\nonumber\\
-\left[\frac{\Gamma_{12}^{++}}{2}(\sigma_1^+a\sigma_2^+a\rho_s-2\sigma_2^+a\rho_s\sigma_1a+\rho_s\sigma_1^+a\sigma_2^+a)\right.\nonumber\\
\left.+\frac{\Gamma_{12}^{--}}{2}(a^{\dag}\sigma_1^-a^{\dag}\sigma_2^-\rho_s-2a^{\dag}\sigma_2^-\rho_sa^{\dag}\sigma_1
+\rho_sa^{\dag}\sigma_1^-a^{\dag}\sigma_2^-)\right.\nonumber\\
\left.+\frac{\Gamma_{12}^{+-}}{2}(\sigma_1^+aa^{\dag}\sigma_2^-\rho_s-2a^{\dag}\sigma_2^-\rho_s\sigma_1^+a
+\rho_s\sigma_1^+aa^{\dag}\sigma_2^-)\right.\nonumber\\
\left.+\frac{\Gamma_{12}^{-+}}{2}(a^{\dag}\sigma_1^-\sigma_2^+a\rho_s-2\sigma_2^+a\rho_sa^{\dag}\sigma_1^-
+\rho_sa^{\dag}\sigma_1^-\sigma_2^+a)\right.\nonumber\\
\left.+1\leftrightarrow2\right],
 \label{appmeq}
\end{eqnarray}
where the first term corresponds to the effective dynamics of the system when QDs are far off-resonant. The effective Hamiltonian
is given by
\begin{eqnarray}
 H_{eff}=H_s+\hbar\sum_{i=1,2}(\delta_i^+a^{\dag}\sigma_i^-\sigma_i^+a+\delta_i^-\sigma_i^+aa^{\dag}\sigma_i^-)\nonumber\\
 -(i\hbar\Omega_{2ph}\sigma_1^-\sigma_2^-a^{\dag 2}+H.c.)\nonumber\\
 -(i\hbar\Omega_{+}\sigma_1^+aa^{\dag}\sigma_2^-+i\hbar\Omega_{-}a^{\dag}\sigma_1^-\sigma_2^+a+H.c.)
\end{eqnarray}
where $\delta_i^{\pm}$ are Stark shifts, the third term represents two-photon processes and the forth term is corresponding to excitation
transfer processes from one QD to another. The expressions for Stark shifts, two-photon transition couplings, and excitation transfer couplings are given by
\begin{eqnarray}
 \delta_i^{\pm}=g_i^2\Im\left[\int_0^{\infty}d\tau G_+e^{\pm i\Delta_i\tau}\right]\\
 \Omega_{2ph}=\frac{g_1g_2}{2}\int_0^{\infty}d\tau(G_--G_-^*)(e^{i\Delta_1\tau}+e^{i\Delta_2\tau})\\
 \Omega_{\pm}=\frac{g_1g_2}{2}\int_0^{\infty}d\tau(G_+e^{\mp i\Delta_2\tau}-G_+^*e^{\pm i\Delta_1\tau}),
 \label{coupl}
\end{eqnarray}
with $G_{\pm}=\langle B\rangle^2(e^{\pm\phi(\tau)}-1)$. The phonon induced cavity mode feeding rates
 $\Gamma_i^{\pm}$, two-photon emission and absorption rates $\Gamma_{ij}^{++}$ and $\Gamma_{ij}^{--}$, and the excitation transfer
 rates $\Gamma_{ij}^{\pm\mp}$ are given by
 \begin{eqnarray}
 \Gamma_i^{\pm}=g_i^2\int_0^{\infty} d\tau(G_+e^{\pm i\Delta_i\tau}+G_+^*e^{\mp i\Delta_i\tau})\\
  \Gamma_{ij}^{++}=g_ig_j\int_0^{\infty} d\tau(G_-e^{i\Delta_j\tau}+G_-^*e^{i\Delta_i\tau})\\
  \Gamma_{ij}^{--}=g_ig_j\int_0^{\infty} d\tau(G_-e^{-i\Delta_j\tau}+G_-^*e^{-i\Delta_i\tau})\\
  \Gamma_{ij}^{+-}=g_ig_j\int_0^{\infty} d\tau(G_+e^{-i\Delta_j\tau}+G_+^*e^{i\Delta_i\tau})\\
  \Gamma_{ij}^{-+}=g_ig_j\int_0^{\infty} d\tau(G_+e^{i\Delta_j\tau}+G_+^*e^{-i\Delta_i\tau}).
 \end{eqnarray}
We solve master equation (\ref{meq}) numerically using quantum optics tool box\cite{toolbox}. In the case when QDs are far off-resonant, the numerical
results by using approximated master equation (\ref{appmeq}) and the results obtained after integration of master equation (\ref{meq})
match perfectly.
\section{Results and Discussions}
For dominating two photon cooperative emission, we consider QDs are off-resonantly coupled with cavity mode.
In Figs.1 to 4, we fix the detuning of one QD, say $\Delta_1$, and
scan the detuning of the other QD for two-photon resonant emission. For subplots (a), (b), (c) and (d), we consider no coupling with phonon bath, coupling with phonon bath at $T=5K$, coupling with phonon bath at $T=10K$, and coupling with phonon bath at $T=20K$, respectively.
We plot photon emission probabilities from state
$|e_1,g_2,1\rangle$, $|g_1,e_2,1\rangle$, and $|g_1,g_2,2\rangle$, given by
$P=\kappa\int_0^{\infty}dt\langle g_1,e_2,1|\rho_s(t)|g_1,e_2,1\rangle$,
$Q=\kappa\int_0^{\infty}dt\langle e_1,g_2,1|\rho_s(t)|e_1,g_2,1\rangle$, and
$R=2\kappa\int_0^{\infty}dt\langle g_1,g_2,2|\rho_s(t)|g_1,g_2,2\rangle$, respectively.  It is clear, that even $g_1$ and $\Delta_1$ are fixed, the probabilities $P$, and $R$ also depend
on $\Delta_2$, which demonstrate that QDs get coupled after interaction with common cavity field and phonon bath. Further, for small spontaneous decay rates $P+Q+R> 0.8$ for $|\Delta_2|\leq5g_1$.
In Fig.1 and Fig.2, we consider that the QDs are placed in the cavity such that they have different dipole coupling constants $g_1\ne g_2$. In Fig.1(a), when there is no coupling with phonon bath and the detuning for first QD is fixed for negative value $\Delta_1=-5g_1$, the probability $P$ remains small and the probability $Q$ becomes maximum for $\Delta_2=0$. The probability $Q$ shows a dip whereas the probability $R$ shows cavity induced two-photon resonance for $\Delta_1+\Delta_2+2g_1^2/\Delta_1+2g_2^2/\Delta_2\approx0$\cite{pathak}, for $g_2=2g_1$ and $\kappa=0.1g_1$, $\Delta_2=2.65g_1$. Further, small values of cavity damping is necessary in order to achieve two photon processes. The appearance of two photon resonance in $R$ is consequence of constructive interference between two photon transitions $|e_1,e_2,0\rangle\rightarrow|e_1,g_2,1\rangle\rightarrow|g_1,g_2,2\rangle$ and $|e_1,e_2,0\rangle\rightarrow|g_1,e_2,1\rangle\rightarrow|g_1,g_2,2\rangle$\cite{pathak}. The cavity induced two-photon resonance satisfies energy conservation $\Delta_1+\Delta_2\approx0$ when we include Stark shifts. In Fig.1(b), we include coupling with phonon bath at $T=5K$. The coupling with phonon reduces the interference between two possible photon transitions thus reduces the probability $R$ at cavity induced two-photon resonance. When QDs are far off-resonant the phonon induced cavity mode feeding enhances single photon processes, thus $P$ and $Q$ increases. The probability $P$ when photon is leaked from state $|g_1,e_2,1\rangle$ and the probability $Q$ when the photon is leaked from state $|e_1,g_2,1\rangle$ complement each other. When probability $P$ increases $Q$ decreases and viceversa. When excitons do not decay through single photon processes, i.e. photon does not emit from state $|e_1,g_2,1\rangle$ or $|g_1,e_2,1\rangle$, two photons are generated in cavity mode and the state of the system is given by $|g_1,g_2,2\rangle$. Therefore, when $R$ increases $P$ and $Q$ decreases. In Fig.1(c) and (d), when temperature of phonon bath increases two-photon processes increase leading to larger probability $R$ for all values of $\Delta_2$. The coupling with phonon bath also open up new phonon induced two-photon resonance when $\Delta_1+2g_1^2/\Delta_1\approx\Delta_2+2g_2^2/\Delta_2$. In this case the two photon transitions $e_1,e_2,0\rangle\rightarrow|e_1,g_2,1\rangle\rightarrow|g_1,g_2,2\rangle$ and $e_1,e_2,0\rangle\rightarrow|g_1,e_2,1\rangle\rightarrow|g_1,g_2,2\rangle$ becomes indistinguishable and interfere constructively again.
\label{Sec:identical}
\begin{figure}[h!]
\centering
\includegraphics[width=6cm, height=6cm]{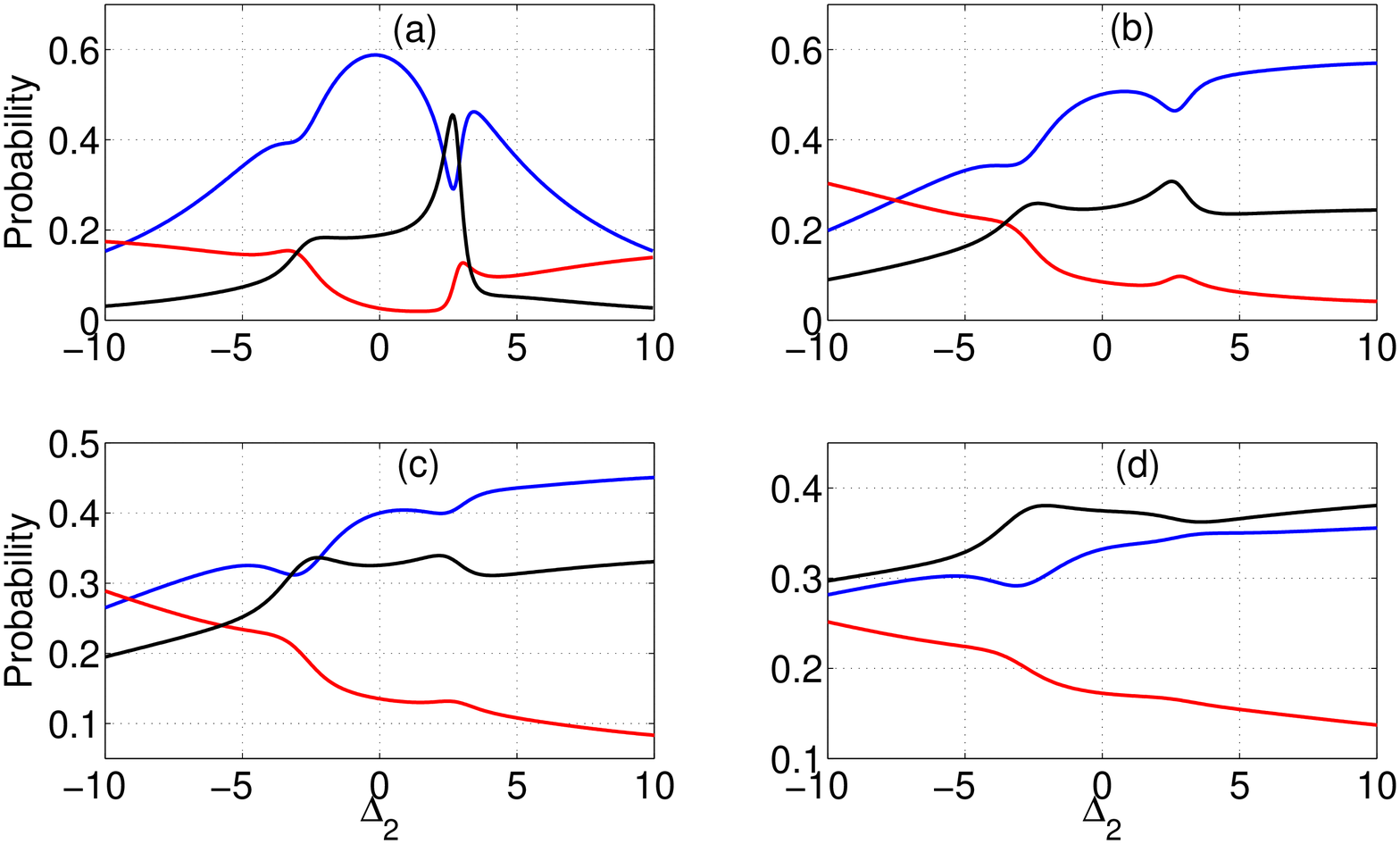}
\caption{The probabilities of photon emission, P from state $|g_1,e_2,1\rangle$ (red line), Q from state $|e_1,g_2,1\rangle$ (blue line), R from state $|g_1,g_2,2\rangle$ (black line). The parameters are $g_2=2g_1$, $\Delta_1=-5g_1$, $\kappa=0.1g_1$, $\gamma_1=\gamma_2=\gamma_1^{\prime}=\gamma_2^{\prime}=0.01g_1$.}
\label{fig1}
\end{figure}

In Fig.2, we fix the detuning of the first QD to positive value $\Delta_1=5g_1$. In this case the cavity induced two-photon resonance appear for $\Delta_2=-2.6g_1$ (see Fig.2(a)). The probabilities $P$, $Q$, and $R$ have same values as in Fig.1(a) but for negative values of $\Delta_2$. In Fig.2(b), when coupling with phonon bath at $T=5K$ is introduced, a prominent phonon induced two-photon resonance appears for $\Delta_1+2g_1^2/\Delta_1\approx\Delta_2+2g_2^2/\Delta_2$. The cavity induced two-photon resonance which appears at $\Delta_2=-2.6g_1$, without coupling with phonon bath, disappears. The probabilities corresponding to single photon processes $P$ and $Q$ increases when QDs are far off-resonant. From fig.1 and Fig.2, it is clear that cavity interaction remains symmetric for positive and negative values of detuning but phonon interaction is asymmetric. The asymmetric behavior of phonon interaction has been observed in QD-cavity systems\cite{meq,asymmetric} earlier.
\begin{figure}[h!]
\centering
\includegraphics[width=6cm, height=6cm]{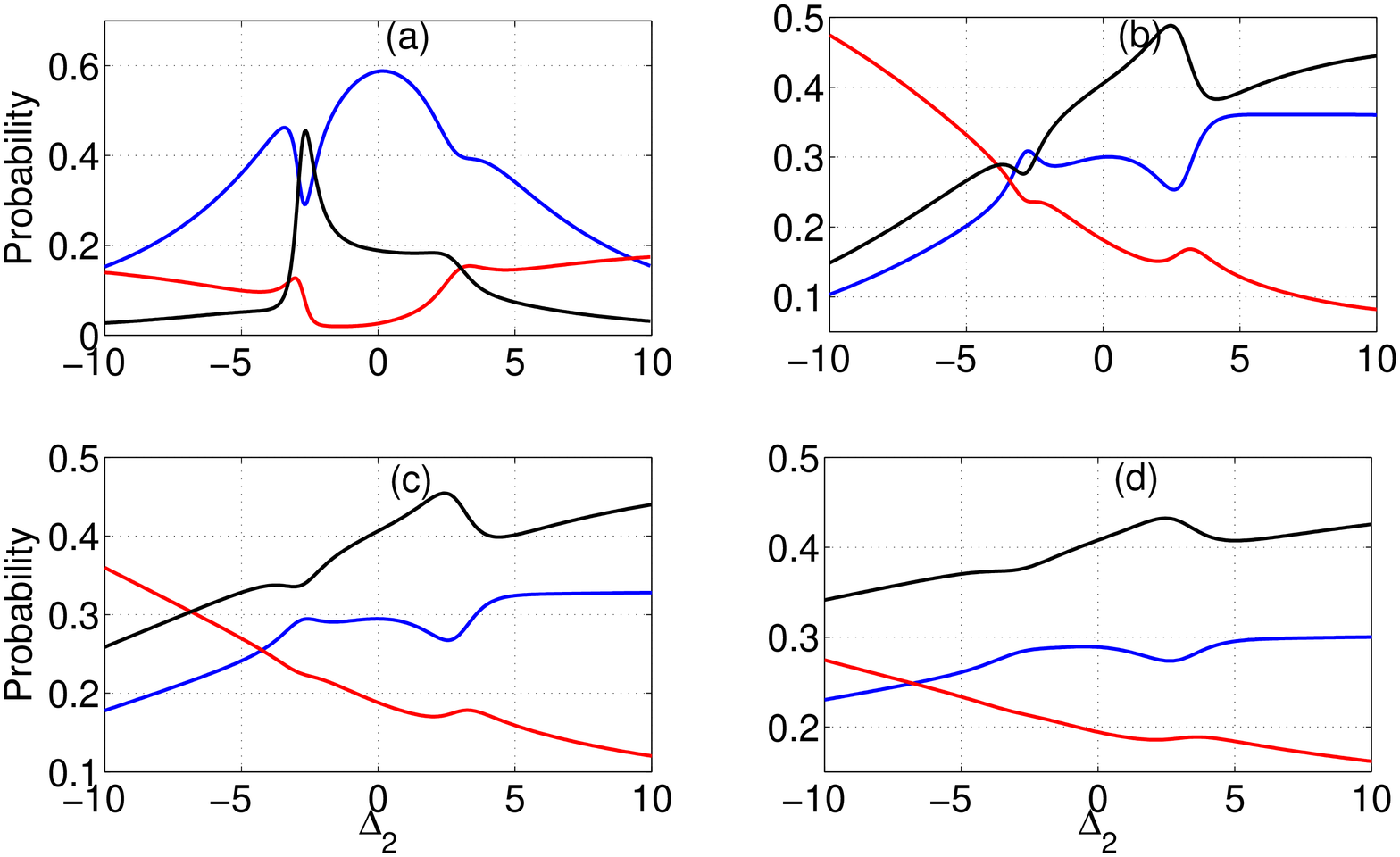}
\caption{The probabilities of photon emission, P from state $|g_1,e_2,1\rangle$ (red line), Q from state $|e_1,g_2,1\rangle$ (blue line), R from state $|g_1,g_2,2\rangle$ (black line). The parameters are same as in Fig.1, except $\Delta_1=5g_1$.}
\label{fig2}
\end{figure}
In Fig.2 (c) to (d), when phonon bath temperature is increased from $T=10K$ to $T=20K$, two-photon processes become larger for all values of $\Delta_2$ except at resonance. At two-photon resonance the probability $R$ decreases slightly as $P$ and $Q$ increase slightly around resonance.
\begin{figure}[h!]
\centering
\includegraphics[width=6cm, height=6 cm]{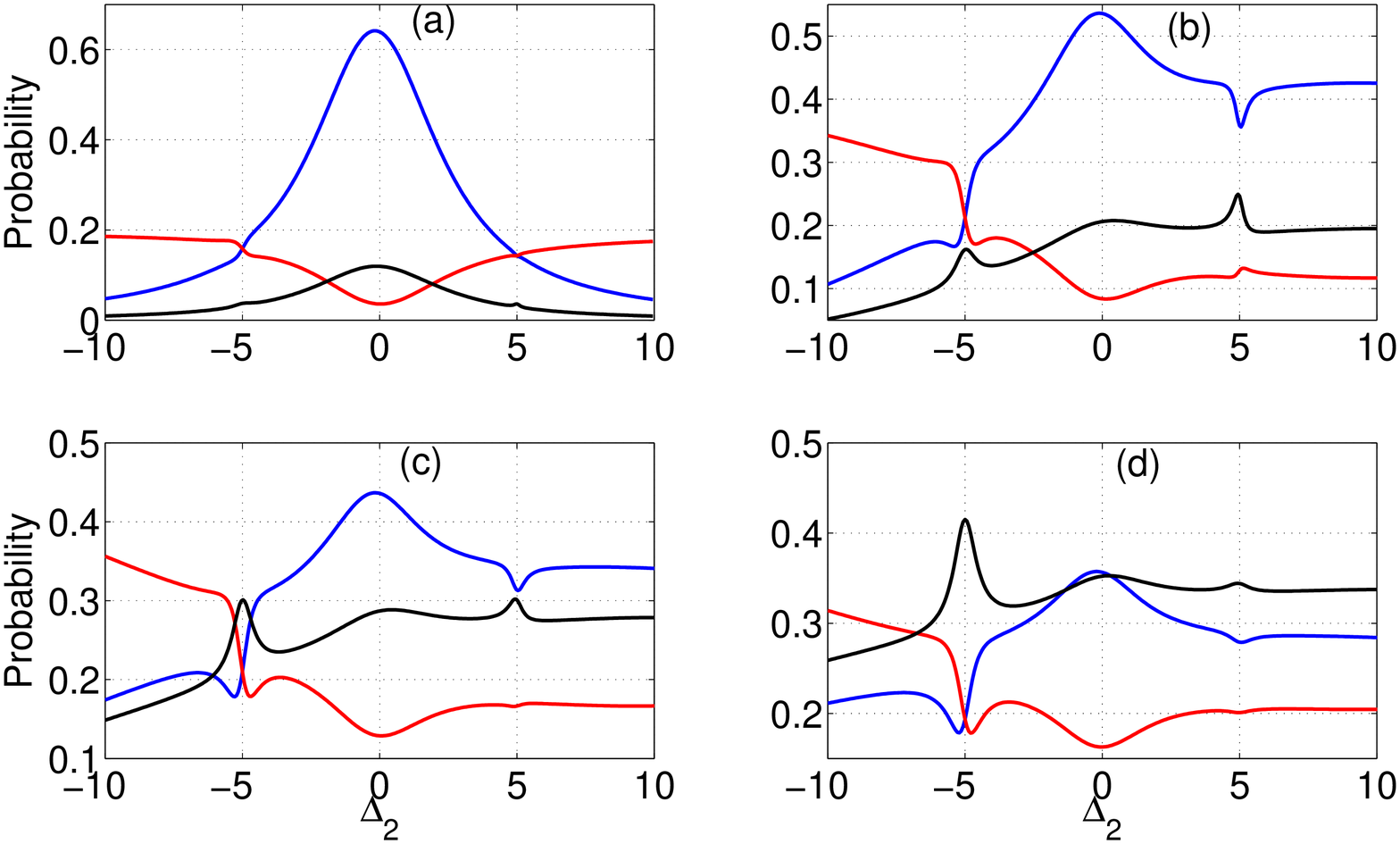}
\caption{The probabilities of photon emission, P from state $|g_1,e_2,1\rangle$ (red line), Q from state $|e_1,g_2,1\rangle$ (blue line), R from state $|g_1,g_2,2\rangle$ (black line). The parameters are $g_2=g_1$, $\Delta_1=-5g_1$, $\kappa=0.1g_1$, $\gamma_1=\gamma_2=\gamma_1^{\prime}=\gamma_2^{\prime}=0.01g_1$.}
\label{fig3}
\end{figure}

In Fig.3 and Fig.4, we consider both QDs have same dipole couplings, $g_1=g_2$. In this case cavity induced two-photon transitions remain negligible\cite{pathak} and single photon transitions dominate, when we do not consider coupling with phonon bath as shown in Fig.3(a) and Fig.4(a). In fact for $g_1=g_2$, the two possible two-photon transitions interfere destructively making the probability of generating state $|g_1,g_2,2\rangle$ negligible. In Fig.3, we fix detuning of the first QD to negative value $\Delta_1=-5g_1$. When we consider electron-phonon coupling at temperature $T=5K$, two-photon processes increase in Fig.3(b). For positive values of $\Delta_2$ two-photon processes are larger than for negative values of $\Delta_2$. Further two tiny peaks appear for $\Delta_2=\pm\Delta_1$. The peak at $\Delta_2=\Delta_1$ corresponds to phonon induced two-photon resonance and the peak at $\Delta_2=-\Delta_1$ corresponds to cavity induced two-photon resonance as the interference conditions change after coupling with phonon bath. In Fig.3(c) and (d), two-photon processes become more dominating leading to larger values of $R$ and smaller values of $P$ and $Q$. At $T=20K$ two photon processes dominate for positive values of $\Delta_2$ with a dominating resonance for negative value at $\Delta_2=\Delta_1$. In Fig.4, we fix $\Delta_1=5g_1$. In Fig.4(b), when electron phonon coupling at $5$K is considered, the two-photon processes increase and dominates over single photon processes leading to larger values of $R$ than $P$ and $Q$ for positive $\Delta_2$. On increasing the temperature, in Fig.4(c) and (d), the two-photon processes become larger and single photon processes decrease. A dominating phonon induced two-photon resonance appears at $\Delta_2=5g_1$.
\begin{figure}[h!]
\centering
\includegraphics[width=6cm, height=6 cm]{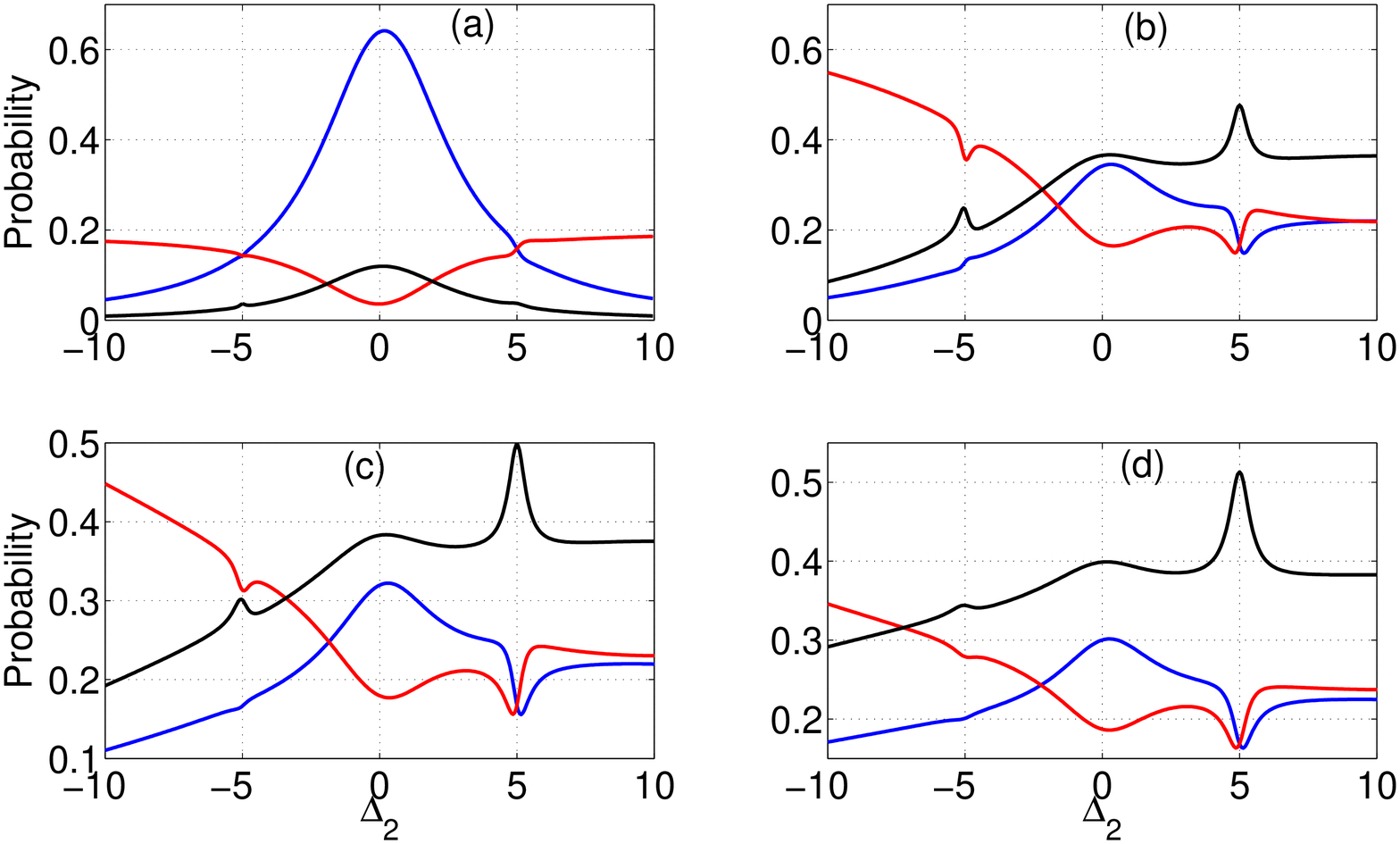}
\caption{The probabilities of photon emission, P from state $|g_1,e_2,1\rangle$ (red line), Q from state $|e_1,g_2,1\rangle$ (blue line), R from state $|g_1,g_2,2\rangle$ (black line). The parameters are same as in Fig.3, except $\Delta_1=5g_1$.}
\label{fig4}
\end{figure}
\begin{figure}[h!]
\centering
\includegraphics[width=6cm, height=6 cm]{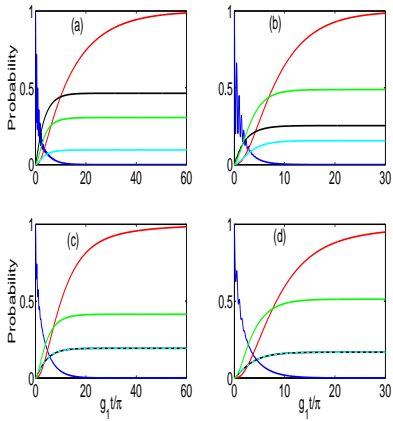}
\caption{The density matrix element $\rho_{ee}(t)=\langle e_1,e_2,0|\rho_s(t)|e_1,e_2,0\rangle$ (blue line), and the probabilities of photon emission, P(t) from state $|g_1,e_2,1\rangle$ (cyan line), Q from state $|e_1,g_2,1\rangle$ (black line), R from state $|g_1,g_2,2\rangle$ (green line) and $R^{\prime}(t)$ from state $|g_1,g_2,1\rangle$ (red line) for $T=10K$. The parameter are, in (a) $g_2=2g_1$, $\Delta_1=-5g_1$, $\Delta_2=2.6g_1$, in (b) $g_2=2g_1$, $\Delta_1=5g_1$, $\Delta_2=2.4g_1$, in (c) $g_2=g_1$, $\Delta_1=\Delta_2=-5g_1$, and in (d) $g_2=g_1$, $\Delta_1=\Delta_2=5g_1$. Other parameters are same.}
\label{fig5}
\end{figure}

In Fig.5, we plot evolution of the system after including electron-phonon interaction at $T=10K$. We plot the probability for both QDs are in excited state $\rho_{ee}(t)=\langle e_1,e_2,0|\rho_s(t)|e_1,e_2,0\rangle$, single photon emission probabilities $P(t)=\kappa\int_0^{t}d\tau\langle g_1,e_2,1|\rho_s(\tau)|g_1,e_2,1\rangle$,
$Q(t)=\kappa\int_0^{t}d\tau\langle e_1,g_2,1|\rho_s(\tau)|e_1,g_2,1\rangle$, and
$R(t)=2\kappa\int_0^{t}d\tau\langle g_1,g_2,2|\rho_s(\tau)|g_1,g_2,2\rangle$. We also plot the probability $R^{\prime}(t)=\kappa\int_0^{t}d\tau\langle g_1,g_2,1|\rho_s(\tau)|g_1,g_2,1\rangle$, when photon is leaked from state $|g_1,g_2,1\rangle$. For smaller values of spontaneous decay rate we find that $R^{\prime}(t)=P(t)+Q(t)+R(t)$. First there is sharp rise in $R^{\prime}(t)$ when population in $|g_1,g_2,1\rangle$ increases due to single photon leakage from $|g_1,g_2,2\rangle$ and then there is slow exponential growth to its maximum value when the transition $|e_1,g_2,0\rangle\rightarrow|g_1,g_2,1\rangle$ and $|g_1,e_2,0\rangle\rightarrow|g_1,g_2,1\rangle$  take place. The probability $\rho_{ee}$ follows rapid oscillations, for smaller values of cavity damping, the average value decays exponentially. In Fig.5(a), we choose $g_2=2g_1$, $\Delta_1=-5g_1$, and $\Delta_2=2.6g_1$, the value of $R(t)$ remains smaller than $Q(t)$. For $g_2=2g_1$, $\Delta_1=5g_1$, and $\Delta_2=2.4g_1$, the probability $R(t)$ dominates, which shows that the phonon induced cooperative two-photon transition from state $|e_1,e_2,0\rangle$ dominates over individual single photon transitions. In (c) and (d), $g_1=g_2$, we notice that for far off-resonant and equally detuned excitons cooperative two-photon decay is always dominating. Further we find that cooperative decay is more pronounced for positive detuning $\Delta_1=\Delta_2=5g_1$ (see Fig.5(d)) than for negative detuning (see Fig.5(c)).
\begin{figure}[h!]
\centering
\includegraphics[width=6cm, height=6 cm]{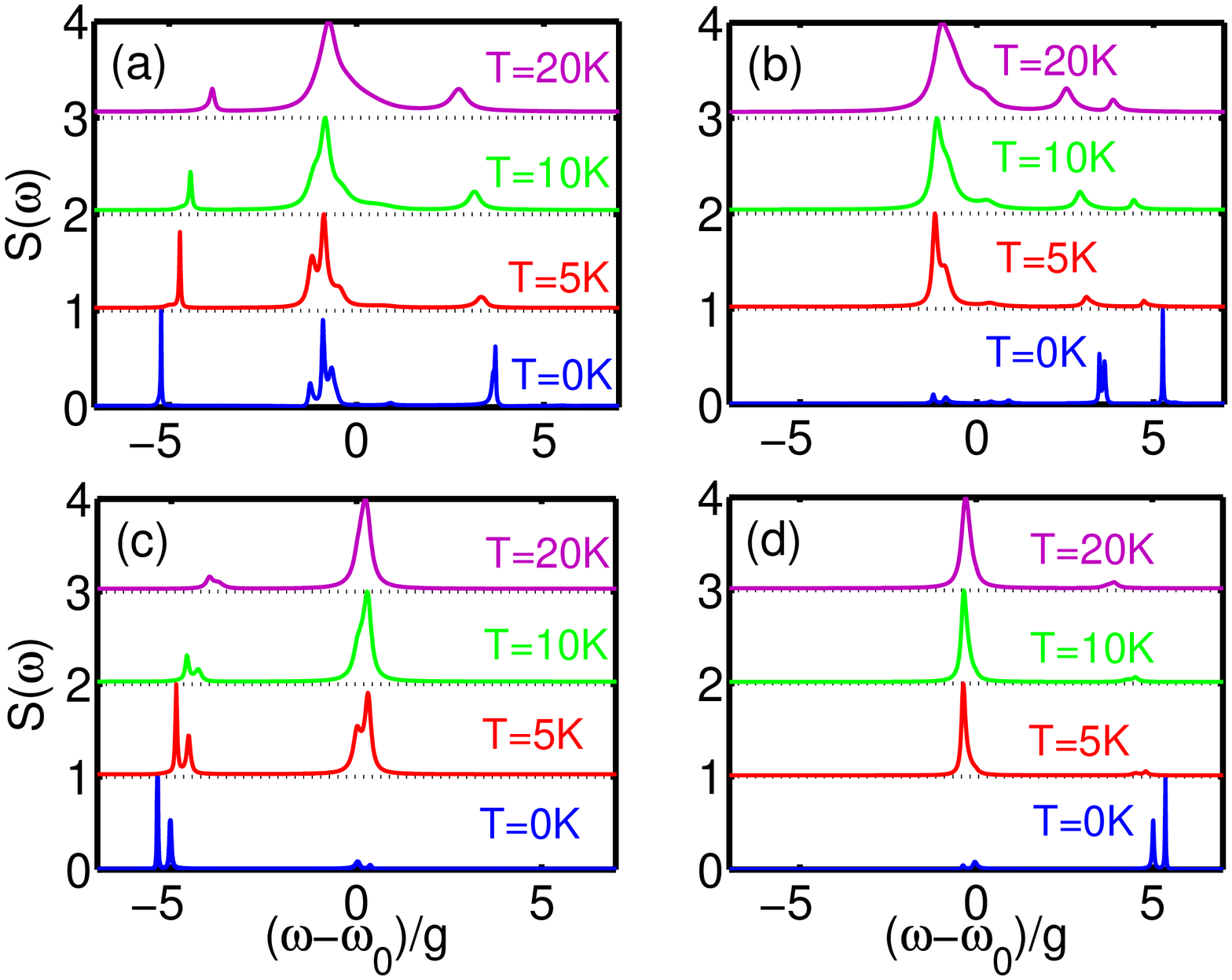}
\caption{The spectrum of photons emitted from cavity mode for parameters same as in Fig.5 but for $T=0K$, $T=5K$, $T=10K$, and $T=20K$.}
\label{fig6}
\end{figure}

In Fig.6, we present spectrum of the emitted photons from cavity mode using similar parameters used in Fig.5 at different phonon bath temperatures. In order to accommodate four subplots for $T=0K$, $T=5K$, $T=10K$ and $T=20K$, we normalize the maximum peak height to $1$ by dividing all values with maximum value. The QDs are coupled with cavity mode under strong coupling regime, therefore spectrum of emitted photons from individual QD has doublet corresponding to exciton like and cavity like frequency\cite{hennessy,doublet}. Further, for far off-resonant QD the emission close to cavity mode remains smaller. When temperature of phonon bath is raised the phonon induced cavity mode feeding becomes larger for single photon transitions leading to emission at cavity frequency. In Fig.6(a), when there is no electron-phonon coupling two exciton like peaks at $\omega-\omega_c\approx-5g_1$ and $\omega-\omega_c\approx2.6g_1$. A cooperative two-photon resonance peak appears around cavity mode frequency which overlaps with two cavity like peaks from individual QDs. When temperature of phonon bath is raised cavity mode feeding increases leading to decrease in emission around exciton frequencies and emission around cavity frequency increases. Further, the peak corresponding to cooperative two-photon emission also start increasing. As a result emission from both single photon processes and cooperative two-photon processes appear around the cavity mode frequency. In Figs.6 (b), (c), and (d), the cavity induced two-photon processes are weak, and we get negligible emission around cavity frequency at $T=0K$. When temperature is raised the emission around cavity frequency dominates due to increase in cavity mode feeding and phonon induced two-photon processes.
\section{Conclusions}
We have predicted dominating two-photon emission from two off-resonantly coupled QDs in a photonic crystal cavity. We have found that when electron-phonon coupling is negligible cavity induced two-photon transition could be dominating over single photon transition if QDs are placed in the cavity such that their dipole coupling constant with cavity mode are not equal ($g_1\neq g_2$) and their exciton transition frequencies satisfy resonant condition $\Delta_1+\Delta_2+2g_1^2/\Delta_1+2g_2^2/\Delta_2\approx0$. For QDs having same dipole coupling constants ($g_1=g_2$), cavity induced two-photon transitions are negligible. In the presence of electron-phonon coupling the cavity induced two-photon transitions are strongly inhibited. However, phonon induced two-photon transitions start dominating with a resonance for $\Delta_1+2g_1^2/\Delta_1\approx\Delta_2+2g_2^2/\Delta_2$. On increasing temperature from $5K$ to $20K$, phonon induced two-photon transitions increase and for the red-detuned cavity mode phonon induced two-photon transitions start dominating at lower temperature than for the blue-detuned cavity mode. Our results can be used for realization of photonic systems when two or more QDs are integrated with micro cavity or waveguide.

\section{Acknowledgements}
This work was supported by DST SERB Fast track young scientist scheme SR/FTP/PS-122/2011.

\end{document}